\documentclass[letter]{aa} 

\usepackage{longtable}
\usepackage{graphicx}
\usepackage{txfonts}
\begin{document}

   \title{Detecting prolonged activity minima in binary stars\thanks{Based on observations made with ESO Telescopes at the La Silla Paranal Observatory under programmes resumed in Table \ref{tabone}, which is only available in electronic form
at the CDS via anonymous ftp to cdsarc.u-strasbg.fr (130.79.128.5)
or via http://cdsweb.u-strasbg.fr/cgi-bin/qcat?J/A+A/.}}
   \subtitle{The case of $\zeta^{2}$ Ret}

\author{M. Flores\thanks{Based on data obtained at Complejo Astronómico El Leoncito, operated under agreement between the Consejo Nacional de Investigaciones Científicas y Técnicas de la República Argentina and the National Universities of La Plata, Córdoba and San Juan. The REOSC observational data and the corresponding reduced spectra (FITS files) are available in electronic form.}
\inst{1,3,5} 
    \and
M. Jaque Arancibia\inst{6}
\and
R. V. Ibañez Bustos\inst{2,4,5}
\and
A. P. Buccino\inst{2,4,5}
\and
J. Yana Galarza\inst{7}
\and
N. E. Nuñez\inst{1,3,5}
\and
P. Miquelarena\inst{1,3,5}
\and
J. Alacoria\inst{1,5}
\and
C. Saffe\inst{1,3,5}
\and
Pablo J. D. Mauas\inst{2,4,5}
}

\institute{Instituto de Ciencias Astron\'omicas, de la Tierra y del Espacio (ICATE), Espa\~na Sur 1512,
    CC 49, 5400 San Juan, Argentina.
    \email{matiasflorestrivigno@conicet.gov.ar}
    \and
Instituto de Astronom\'ia y F\'isica del Espacio (IAFE), Buenos Aires, Argentina.
\and
Facultad de Ciencias Exactas, F\'isicas y Naturales, Universidad Nacional de San Juan, San Juan, Argentina.
\and
Departamento de F\'isica, Facultad de Ciencias Exactas y Naturales, Universidad de Buenos Aires, Buenos Aires, Argentina.
\and
Consejo Nacional de Investigaciones Cient\'ificas y T\'ecnicas (CONICET), Argentina.
\and
Departamento de F\'isica y Astronom\'ia, Universidad de La Serena, Av. Cisternas 1200, La Serena, Chile
\and
Universidade de S\~ao Paulo, Departamento de Astronomia do IAG/USP, Rua do Mat\~ao 1226, \
     Cidade Universit\'aria, 05508-900 S\~ao Paulo, SP, Brazil.}

\date{December 2020}

\abstract
 {It is well known that from 1645 to 1715 solar activity was notably low and the number of sunspots was extremely reduced. This epoch is known as the Maunder Minimum (MM). 
The study of stars in prolonged activity minima like the MM could help to shed light on this enigmatic epoch. However, to date, it is not easy to identify
MM candidates among other stars. An original idea, that has been little explored, is to compare the activity levels of both components of binary systems.}       
{To explore if the star $\zeta^{2}$ Ret, which belongs to a binary system, is in (or going to) a state similar to the MM. We have collected more than 430 spectra acquired between 2000 and 2019 with the HARPS, REOSC, UVES, and FEROS spectrographs.}
{We performed a detailed long-term activity study of both components using the Mount Wilson index, which is obtained from the \ion{Ca}{ii} H\&K lines. To search for signs of an activity cycle, we analyzed the resulting time-series with the Generalized Lomb-Scargle and CLEAN periodograms.}
{Our spectroscopic analysis shows a high activity level for $\zeta^{1}$ Ret and a significant decrease in the magnetic activity cycle amplitude of $\zeta^{2}$ Ret. Thus, the activity difference between both components has been slightly increased ($\Delta$log ${R}'_\mathrm{HK}\sim$0.24 dex), when compared to the previously reported value. The long series analyzed here allowed us to recalculate and constrain the period of $\zeta^{2}$ Ret to $\sim$ 7.9 yr. We also detected a long-term activity cycle of $\sim$4.2 yr in $\zeta^{1}$ Ret, which has not yet been reported in the literature.} 
{By analogy with the scenario that proposes a weak solar cycle during the MM, we suggest that activity signatures showed by $\zeta^{2}$ Ret, i.e., a very low activity level when compared to its stellar companion, a notably decreasing amplitude ($\sim$47\%), and a cyclic behaviour, are possible evidence that this star could be in a MM state. It is, to our knowledge, the first MM candidate star detected through a highly discrepant activity behaviour in a binary system. Finally, we suggest that continuous observations of $\zeta^{2}$ could help to better understand unusual periods such as the MM.}

   \maketitle

\section{Introduction}

A pioneering research based on the monitoring of H\&K fluxes for 91 main-sequence stars showed that activity variations, including long-term cyclic behaviour similar to the 11-yr cycle of Sun, were also observed in other stars \citep{1978ApJ...226..379W}. Then, by using the Mount Wilson index ($\mathrm{S}_\mathrm{{MW}}$) defined as the ratio between the flux in the optical \ion{Ca}{ii} H\&K lines and the nearby continuum  \citep{1978PASP...90..267V}, \citet{Baliunas98} analyzed a sample of 2200 stars, finding  different types of long-term activity behaviour. Those stars with a cyclic behaviour (with periods between 2.5 and 25 yr) showed intermediate activity levels, while those with erratic behaviour had higher activity levels. A third group presented flat activity levels, in general corresponding to inactive stars. These objects are particularly interesting because they could be in a state similar to the solar Maunder Minimum (hereafter MM).

The MM was a phase between 1645 and 1715 when the Sun deviated from its usual 11-yr activity cycle \citep{1976Sci...192.1189E}, the number of sunspots was extremely reduced, although they did not disappear \citep{1993A&A...276..549R}. In addition, some evidence, including other solar proxies, have suggested that the solar cycle was still in progress although with reduced amplitude during the MM \citep[e.g.][]{1989AnGeo...7..321R,1993A&A...276..549R,1998SoPh..181..237B,2001JGR...10616039U,2003mmvs.book.....S,2007AstL...33..340N,2015A&A...577A..71V,Zolotova_2015}.

In particular, the study of MM analogue stars could be very useful to better understand the Sun’s magnetic field, especially its evolution in the past and future. It is also relevant to improve our knowledge about the current dynamo models \citep[e.g.][]{1992ASPC...27..150S,2006A&A...457L..25U,2010LRSP....7....3C,Shah_2018}. However, the detection of a  MM analogue state is a challenging task due to the long-term monitoring that is required, as well as to the lack of a clear criteria to identify MM candidates \citep[e.g.][]{2004AJ....128.1273W,2007ApJ...663..643J}.

Initial efforts to establish a criterion were carried out by \citet{1990Natur.348..520B} and \citet{1995ApJ...438..269B}, which was based 
on the analysis of the relative variation of $\mathrm{S}_\mathrm{{MW}}$ index around its mean ($\sigma_{\mathrm{S}}$/$\overline{\mathrm{S}}$). In this sense, the authors initially considered as MM candidates those stars with $\sigma_{\mathrm{S}}$/$\overline{\mathrm{S}}$ $<$ 1.5\%\footnote{While those stars with an $\sigma_{\mathrm{S}}$/$\overline{\mathrm{S}}$ $\geq$ 2\% are considered variable or erratic.}. They were called ``flat'' stars and were characterized by relatively constant and low activity levels.
 
Then, \citet[][]{1996AJ....111..439H} studied a sample of stars that belong to the Project Phoenix Survey. As a result, the authors 
define a new class of inactive stars employing the chromospheric activity index log ${R}'_\mathrm{HK}$.  
According to their definition, stars with a log ${R}'_\mathrm{HK}< -5.1$ dex (corresponding to $\mathrm{S}_\mathrm{{MW}}$ $<$ 0.15 for solar type stars) could be considered as MM candidates. However, \citet{2004AJ....128.1273W} showed that most of these stars were in fact evolved stars, with activity levels significantly lower than main-sequence objects, thereby concluding that the low activity alone is not a sufficient discriminant of a MM state. Moreover, it has been suggested that the low activity level obtained by \citet[][]{1996AJ....111..439H} 
should be higher than $-5.1$ dex and that the identification of MM candidates should not be constrained only to 
the visual part of the spectra. In this way, UV and X-ray data can also be used to identify MM candidates \citep[e.g.][]{2004AJ....128.1273W,2007ApJ...663..643J}. 
Recently, \citet{2017MNRAS.470..276S} reported an average S-index of 0.154 for the unusually deep and long minimum in 2008-2009 of the solar cycle 24. The solar far-UV data also reveals a low activity behaviour during this period. Then, according to the analysis of the authors, the Sun could be entering into a new grand minimum phase.
   
To date, there are only very few firm MM candidates reported in the literature. For instance, \citet{2009A&A...508.1417P} 
analyzed the exoplanet host star 51 Pegasi by using both X-ray and \ion{Ca}{ii} H\&K data. A constant and low coronal flux  
in addition to a flat chromospheric activity suggest that this star could be in a MM state. Another example is the star 
HD\,4915, which has been recently reported as a MM candidate by \citet{Shah_2018}. In that work, the authors studied 
the activity behaviour of the star by using a long-term database of the \ion{Ca}{ii} H\&K optical lines (acquired between 2006 and 2018).  
They found a decrease of the magnetic activity over two cycles, revealed by the core flux variation of \ion{Ca}{ii} H\&K lines. This fact could be a strong indication for a possible MM state in HD 4915.

An alternative way to identify MM candidates was proposed by \citet[][]{1998ASPC..154.1235D} (hereafter DO98) and \citet{2004AJ....128.1273W}.
The authors pointed out that a remarkable difference in the activity behaviour among main-sequence binary components could be used as a MM stars detector.
In such systems, a similar MM state could be associated to the star with lower activity. Following this interpretation, \citet{2018MNRAS.476.2751F} suggested that the activity difference, observed between the components of the $\zeta$ Ret binary system, could be attributed to an atypical activity of $\zeta^{2}$ Ret. The F$_X$ values estimated from XMM-Newton database for $\zeta^{1}$ Ret and $\zeta^{2}$ Ret are (5.11 $\pm$ 0.08) $\times 10^{-13}$ and (0.25 $\pm$ 0.32) $\times 10^{-13}$ erg s$^{-1}$ cm$^{-2}$, respectively. This shows that $\zeta^{1}$ Ret is more active than $\zeta^{2}$ Ret in X-rays \citep[see][for more details]{2018MNRAS.476.2751F}. In this way, as a feasible scenario, the star $\zeta^{2}$ Ret is possibly emerging from (or going to) a state similar to the MM. In that work, we stressed the need for additional spectroscopic data in order to verify or rule out this possible scenario. Fortunately, more spectroscopic ESO data were acquired for this remarkable binary system. Moreover, we count with additional spectra taken with the REOSC spectrograph at CASLEO observatory.

This binary system is conformed by two solar analogue stars physically connected \citep{2011ApJS..192....2S}, and their spectral types are classified as G2 V and G1 V according to the Hipparcos database \citep[see][for details]{2016A&A...588A..81S}. Both stars have very similar stellar parameters ($T_{eff}$, log $g$, and [Fe/H]) and they are also similar to the Sun \citep[see][for more details]{2016A&A...588A..81S}. Their empirical rotational periods obtained from the \citet{2008ApJ...687.1264M} calibration are 13.2 $\pm$ 2.8 d and 16.5 $\pm$ 1.8 d for  $\zeta^{1}$ Ret and $\zeta^{2}$ Ret, respectively. This strong physical similarity could help to diminish or remove a possible dependence of the minimum \ion{Ca}{ii} H\&K activity levels on gravity and metallicity \citep[e.g.][]{1989ApJ...341.1035S,2004AJ....128.1273W,2012IAUS..286..257G}, which is an additional advantage for the mutual comparison in this system. Besides, the large available data set ($\sim$ 19 yrs of observations) converts this system in a unique laboratory that allows us to carry out a detailed long-term activity study in order to explore the possible MM state of $\zeta^{2}$ Ret, following the suggestion of DO98.

The paper is organized as follows: in \S 2, the observations and data reduction are described. In \S 3, our stellar activity analysis is presented. Finally, our discussion and main conclusions are exposed in \S 4.

\section{Observations and data reduction}
Most of the stellar spectra of $\zeta^{1}$ ($=$HD\,20766) and $\zeta^{2}$ Ret ($=$HD\,20807) were downloaded from the European Southern Observatory (ESO) archive\footnote{\url{http://archive.eso.org/wdb/wdb/adp/phase3_spectral/form?phase3_collection=HARPS}}. These observations were acquired with the \textrm{HARPS} spectrograph (resolving power R $\sim$ 115\,000), attached to the La Silla 3.6-m (ESO) telescope between 2003 and 2019. We also included some spectra taken with UVES (between 2002 and 2009) and FEROS (during 2010 and 2014) spectrographs (R$\sim$ 80\,000 and R$\sim$48\,000, respectively), which are coupled to the Unit 8.2-m Telescope 2 (UT2) of the Very Large Telescope (VLT) and to the 2.2-m telescope located at La Silla, respectively. All ESO spectra have been automatically processed by the corresponding pipelines\footnote{\url{http://www.eso.org/sci/facilities/lasilla/instruments/harps/overview.html}}$^{,\thinspace}$\footnote{\url{http://www.eso.org/sci/facilities/paranal/instruments/uves.html}}$^{,\thinspace}$\footnote{\url{http://www.eso.org/sci/facilities/lasilla/instruments/feros.html}}.

Additionally, our analysis was complemented with observations performed with the REOSC\footnote{\url{https://casleo.conicet.gov.ar/reosc-ds-dc/}} spectrograph (R$\sim$13\,000), working at the 2.15-m Jorge Sahade telescope at the CASLEO in San Juan, Argentina. These data, taken between 2000 and 2015 under the HK$\alpha$ 
project\footnote{The main aim of the HK$\alpha$ project consists in the systematic observation of main sequence stars to carry out long-term activity studies \citep[see][for more details]{2004A&A...414..699C}}, were reduced following the standard procedures with IRAF\footnote{IRAF is distributed by the National Optical Astronomical Observatories, which is operated by the Association of Universities for Research in Astronomy, Inc. (AURA), under a cooperative agreement with the National Science Foundation.} tasks, i.e. performing bias subtraction, flat fielding, sky subtraction, order extraction, and wavelength calibration. See Table \ref{tabone} for observation logs details.

Before the calculation of the standard $\mathrm{S}_\mathrm{{MW}}$ index  defined by \citet{1978PASP...90..267V} at the Mount Wilson Observatory (MWO), we first discard those spectra with low signal-to-noise ratio (S/N $\leq$ 100). As a result, we obtained 79 spectra for $\zeta^{1}$ Ret and 352 for $\zeta^{2}$ Ret. These spectra, with a mean S/N $\sim$175 at 6070 Å, were corrected by radial velocities using standard IRAF tasks. 
Then, we integrated the flux in two windows centred at the cores of the Ca {\sc ii} H\&K lines (3968.47 {\AA} \ and 3933.66 {\AA}, respectively), weighted with triangular profiles of 1.09 {\AA} full width at half-maximum (FWHM), and computed the ratio of these fluxes to the mean continuum flux, integrated in two passbands of $\sim$ 20 {\AA} width centred at 3891 and 4001 {\AA}. As a result, we obtained the S-index corresponding to each one of the 
instruments used in this work, which are then converted to the $\mathrm{S}_\mathrm{{MW}}$ following the calibration procedures of \citet[][]{2011arXiv1107.5325L}, \citet[][]{2007A&A...469..309C}, and \citet[][]{2008A&A...485..571J} for HARPS, REOSC and FEROS spectroscopic data. 
For the case of UVES spectra, there is no calibration available. These data were intercalibrated to the rest of the time-series.

\section{Stellar Activity Analysis}

\begin{figure}
\centering
\includegraphics[width=\columnwidth]{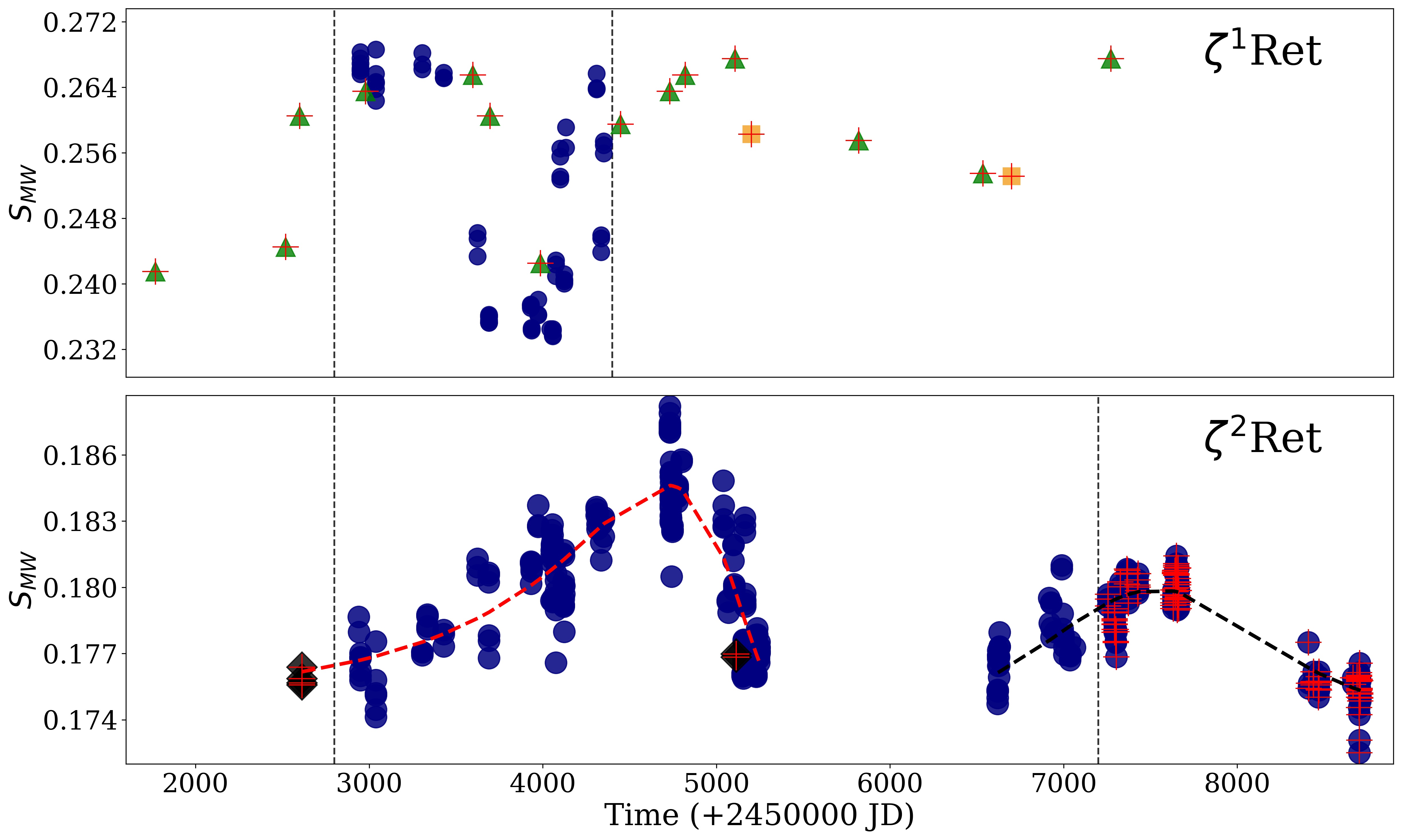}
\caption{Upper panel corresponds to the $\mathrm{S}_\mathrm{{MW}}$ index variation of $\zeta^{1}$ Ret. HARPS data are indicated with blue circles (for both panels), while REOSC and FEROS data are indicated with green triangles and orange squares, respectively. Lower panel: activity variation for $\zeta^{2}$ Ret. Here, UVES data are indicated with black diamonds. Both vertical dashed lines in each panel represent the time coverage of those series reported initially, while the new data are indicated with red crosses. Red and black dashed lines show the fitted activity maxima $f(t)$ for both peaks.}
\label{plot.0}
\end{figure}

In order to search for clear signatures of a possible MM state in the binary system $\zeta$ Ret as suggested in \citet{2018MNRAS.476.2751F}, in Fig. \ref{plot.0}  we show the time-series of the $\mathrm{S}_\mathrm{{MW}}$ indexes for both components ($\zeta^{1}$ Ret and $\zeta^{2}$ Ret are plotted in the upper and lower panels, respectively). We have included all spectroscopic data from HARPS, REOSC, FEROS, and UVES. As a result, we have an extensive database for each component of approximately 19 years. Vertical dashed lines were plotted to highlight the time
coverage of our current series from those published in \citet{2018MNRAS.476.2751F}.

A direct comparison of these new time-series shows a clear decrease in the amplitude of the chromospheric activity of the $\zeta^{2}$ Ret component (from the first peak to the last one). To quantify this decrease of activity, in Fig. \ref{plot.0} we fitted the corresponding time-series assuming a typical solar activity shape $f(t)$ for each peak \citep[see Eq. 8 in][for more details]{2017ApJ...835...25E}. The first cycle fit (red dashed line) is given by the following parameters: $A= 0.0104$, $B=1.32$ yr, $\alpha=-0.24$ yr$^{-2}$, $t_{m}=2008.67$ yr (xJD=4711.72 days), and $f_{min}=0.17422$, while the corresponding parameters to the second cycle (black dashed line) are $A=0.0053$, $B=1.59$ yr, $\alpha\sim 2\times 10^{-26}$ yr$^{-2}$, $t_{m}=2016.42$ yr (xJD=7542.22 days), and $f_{min}=0.1746$. As a result, both fits show an amplitude $\Delta S_{MW}=0.0084$ and $\Delta S_{MW}=0.0045$, meaning a decrease of $\sim$47\% in the activity cycle amplitude. To classify each component according to their variability type, we considered the criteria adopted by \citet{1995ApJ...438..269B}. Then, $\zeta^{1}$ Ret can be classified as a variable star (with $\sigma_{\mathrm{S}}$/$\overline{\mathrm{S}}$ $\sim$ 4.3\%), while $\zeta^{2}$ Ret would be classified as a ``flat star'' ($\sigma_{\mathrm{S}}$/$\overline{\mathrm{S}}$ $\sim$ 1.4\%), although its stellar activity shows a clear variation. For comparative purposes, we also computed the $\log {R}'_\mathrm{HK}$ index by subtracting the photospheric contribution following the prescription given in \citet{1984ApJ...279..763N}, resulting in a mean activity difference of $\sim$0.24 dex, which is slightly higher to the previous value of 0.22 dex reported in \citet{2018MNRAS.476.2751F}.

In order to explore the components of the \ion{Ca}{ii} H\&K line-core fluxes responsible for the low $\mathrm{S}_\mathrm{{MW}}$ index in $\zeta^2$ Ret, we computed its basal level of activity.  \cite{1989ApJ...341.1035S} conclude that the line-core emission in the \ion{Ca}{ii} lines is composed by a photospheric component, a basal flux probably related to acoustic heating and a third component associated to purely magnetic activity. In this sense, we estimated the \ion{Ca}{ii} photospheric component using a synthetic spectra calculated with SYNTHE and ATLAS9 model atmospheres \citep{1993KurCD..13.....K}, accounting for a photospheric Mount Wilson index $S_{Phot}=0.149$. Following \cite{2013A&A...549A.117M}, we converted this index to the \ion{Ca}{ii} line-core  photospheric fluxes of (2.539$\pm$ 0.017)$\times 10^6$ erg cm$^{-2}$ s$^{-1}$, higher than the photospheric flux derived in \cite{1984ApJ...279..763N} for a star of $B-V=0.60$. \citet{2013A&A...549A.117M} revised this historical work and obtained that the photospheric flux was underestimated. They also  computed an excess basal flux non-photospheric in erg cm$^{-2}$ s$^{-1}$ given by $log(F'_{HK})=6.42-1.03(B-V)$. Considering this contribution $F'_{HK}=(5.65\pm 0.51)\times 10^5$ erg cm$^{-2}$ s$^{-1}$ and the photospheric flux derived from $S_{phot}$ for $\zeta^2$ Ret, all the basal contribution $F'_{HK}+F^{phot}_{HK}=(3.10\pm 0.07)\times 10^6$ erg cm$^{-2}$ s$^{-1}$ is associated to  a Mount Wilson index of $S_{MW}\sim 0.180$. The mean activity level of $\zeta^2$ Ret between 6658.5 and 8849.5 days is slightly lower in less than 1.5$\sigma$, thus mainly related to a basal chromospheric heating. Although a remaining magnetic contribution is still evident in the activity cycle.

\begin{figure}
\centering
\includegraphics[width=\columnwidth]{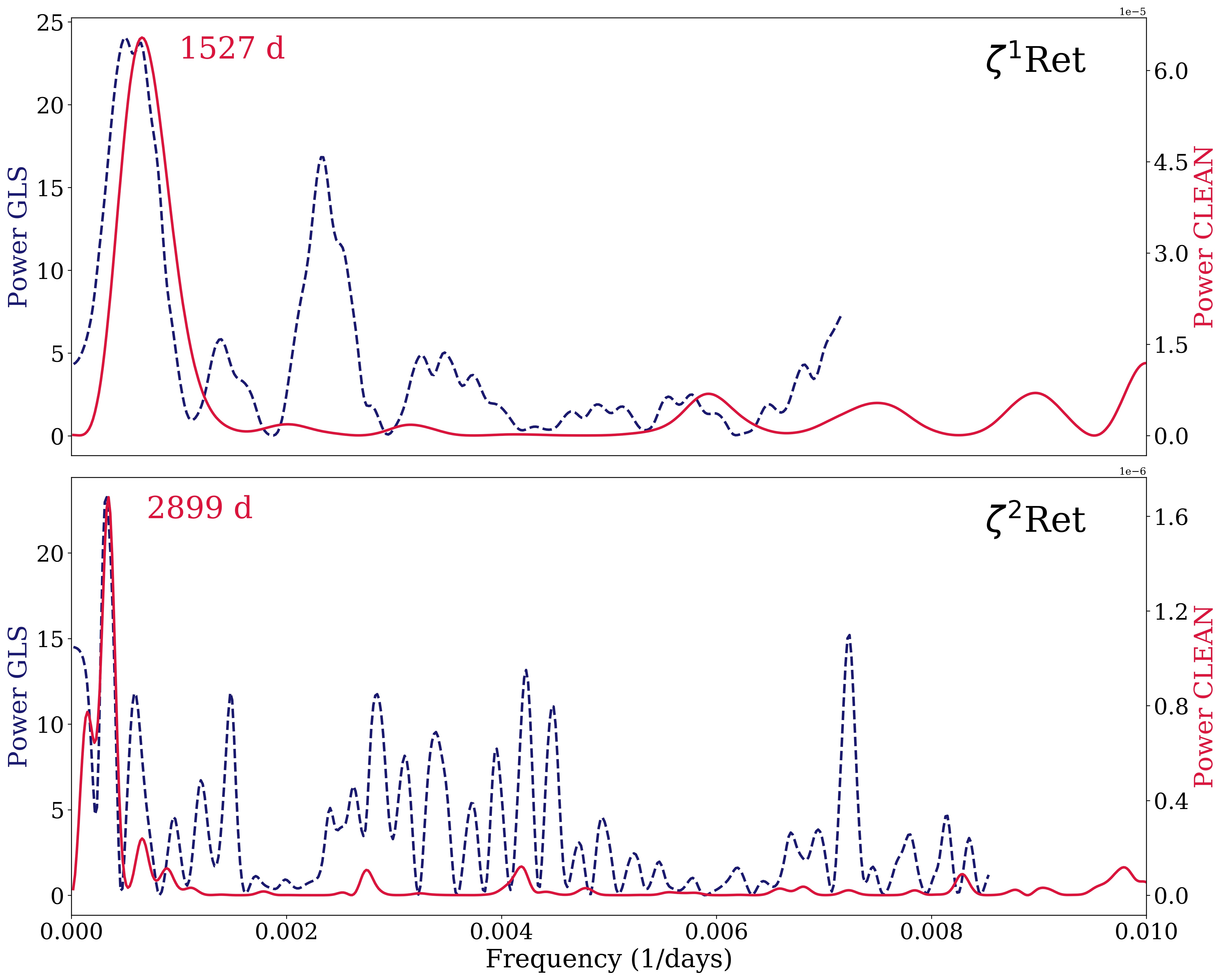}
\caption{GLS (blue dashed line) and CLEAN (red continuous line) periodograms for the means of the Mount Wilson indexes plotted in Fig. \ref{plot.0}. Upper and lower panels correspond to $\zeta^{1}$ Ret and $\zeta^{2}$ Ret, respectively. The most significant CLEAN periods are indicated in each panel.}
\label{plot.2}
\end{figure}

To search for long-term activity cycles in this binary system, we first calculated the monthly means of all data. This procedure, which has been applied in previous works \citep[e.g.][]{1995ApJ...438..269B,2010ApJ...723L.213M,2011A&A...534A..30G,2018A&A...620A..34F}, enables us to reduce the rotational scatter originated by individual active regions. Following \citet{2009A&A...496..577Z}, we computed the Generalized Lomb-Scargle periodogram (hereafter GLS) and the false-alarm probability (hereafter FAP, see their equation 24) of each significant peak present in the periodograms. In the upper and lower panels of Fig. \ref{plot.2} we show the GLS (blue dashed line) for $\zeta^{1}$ Ret and $\zeta^{2}$ Ret, respectively. Both stars seem to be periodic. In the case of $\zeta^{1}$, we found two prominent peaks which can be associated with activity cycles, one of them has a period of 1548 $\pm$ 62 d with a FAP of  1 $\times 10^{-14}$. While, the second peak is 431 $\pm$ 6 d with a  FAP of 2 $\times 10^{-08}$. For the $\zeta^{2}$ Ret component, the large data set collected in this work allowed us to recalculate its previously reported period (3670 $\pm$ 170 d). A prominent peak of 3047 $\pm$ 134 d with a  FAP of 4 $\times 10^{-12}$ was detected in the GLS periodogram (see the lower panel of Fig. \ref{plot.2}). Therefore, $\zeta^{2}$ Ret also satisfies the \citet{1995ApJ...438..269B} criteria for a cycling star (i.e. FAP $\leq 10^{-02}$).

We also executed the CLEAN deconvolution algorithm \citep{Roberts87} to explore whether the 431-day period which appears in the GLS periodogram of $\zeta^{1}$ Ret is due to sampling. A comparison between GLS (blue dashed line) and CLEAN (red continuous line) periodograms is shown in Fig. \ref{plot.2}. For the case of $\zeta^{1}$, we found that the only predominant peak is around 1527 $\pm$ 43 d, while for $\zeta^{2}$ we obtained a single significant period of 2899 $\pm$ 139 d. The errors of the periods detected with the CLEAN algorithm depends on the finite frequency resolution of the periodograms $\delta\nu$ as given by Eq.(2) in \cite{Lamm04}, $\delta P=\frac{\delta\nu P^2}{2}$. 
Therefore, the periods found for the binary system employing the CLEAN algorithm correspond to those associated with the GLS method.

\begin{figure}
\centering
\includegraphics[width=\columnwidth]{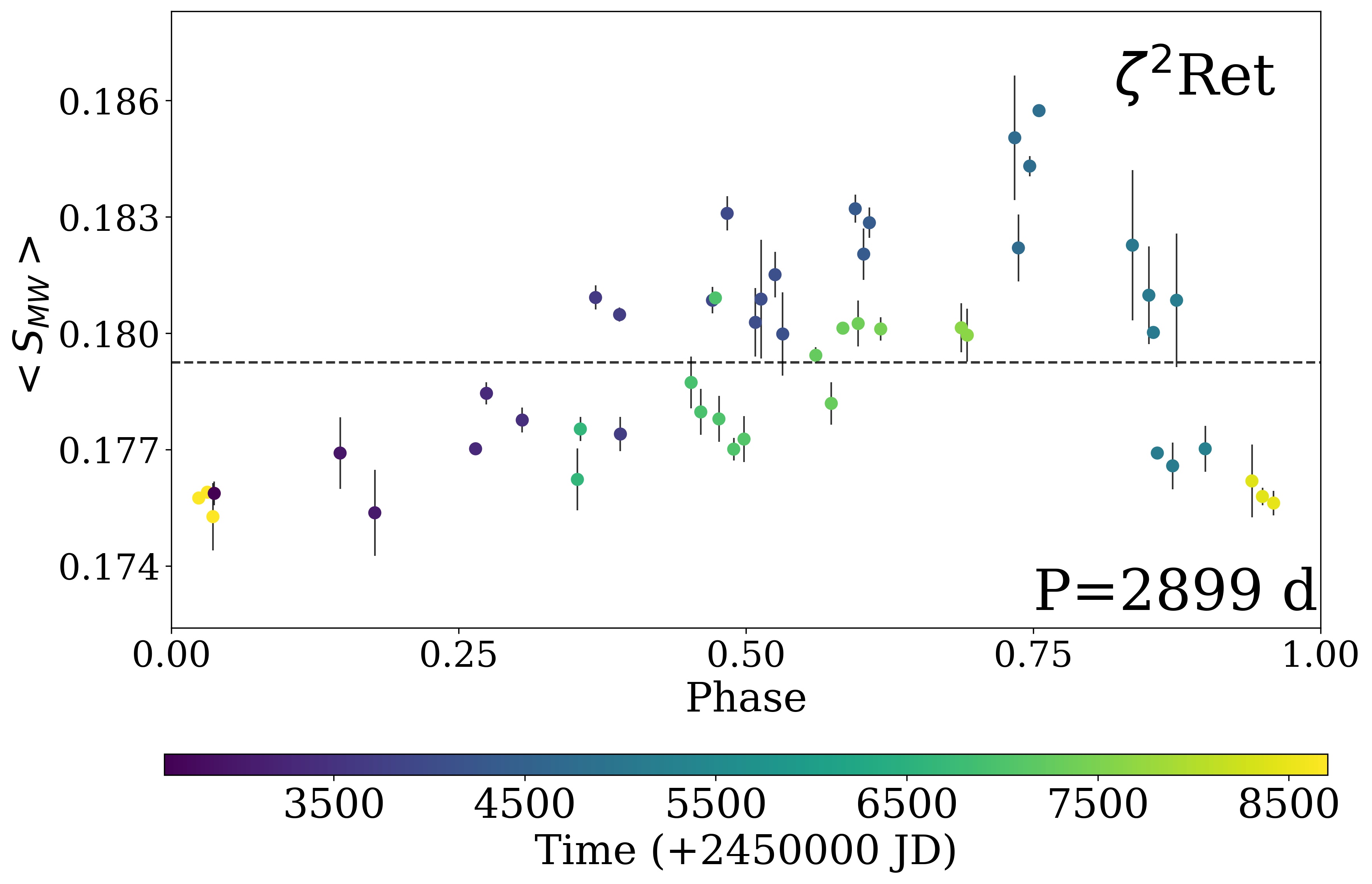}
\caption{Monthly means of the Mount Wilson indexes for $\zeta^{2}$ Ret phased with a period of $\sim$ 7.9 yr. The observing seasons are represented by coloured circles. The error bars of HARPS data and the corresponding mean activity level (dashed horizontal line) have been included.}
\label{plot.3}
\end{figure}

Finally, in Fig. \ref{plot.3} we show the monthly average values of the $\mathrm{S}_\mathrm{{MW}}$ index for $\zeta^{2}$ Ret phased with the period derived with the CLEAN algorithm. The errors of these data have been calculated as their standard deviation, while for those bins with a single measurement we have considered the typical dispersion of other bins.

\section{Discussion and conclusions}

Following the aim of this study, we carried out a long-term activity of the $\zeta$ Ret binary system 
employing several spectroscopic data obtained over a span of 19 years. We detected long-term activity cycles of 1527 $\sim$ 43 d ($\sim$ 4.2 yr, not yet reported in the literature) and 2899 $\sim$ 139 d ($\sim$ 7.9 yr) for $\zeta^{1}$ and $\zeta^{2}$ Ret, respectively. 
In particular, the new data included in this work allowed us to make a better estimation of
the period obtained in \citet{2018MNRAS.476.2751F} for $\zeta^{2}$ Ret.

\citet{2018MNRAS.476.2751F} proposed two possible scenarios to explain
the large difference in activity between the stars $\zeta^{1}$ and $\zeta^{2}$ Ret.
In the first scenario, the stars possibly present different rotational periods\footnote{The analysis of these new spectroscopic data does not reveal the presence of any reliable rotational modulation for both stars.}, which could result in different average activity levels.
The second scenario suggests that the star $\zeta^{2}$ Ret is possibly in a MM state.
In the present work we collected new evidence, including more than 430 spectra
acquired with the HARPS, REOSC, FEROS, and UVES spectrographs, which support the idea that $\zeta^{2}$ Ret is possibly in a MM state, 
due to the following reasons:

\begin{itemize}

\item[--] A large difference in the average activity levels between both stars (hereafter $\Delta$).
DO98 and \citet{2004AJ....128.1273W} suggested that binary systems (presumably co-eval stars),
with significant $\Delta$ levels, could point to a MM state of the most inactive star.
In particular, DO98 suggest that an age difference (estimated using an activity-age
calibration) greater than $\sim$1 Gyr could indicate a MM state. Using the DO98
calibration, we estimate ages of 1.5 and 3.3 Gyr for $\zeta^{1}$ Ret and $\zeta^{2}$ Ret
i.e. a notable difference of $\sim$1.8 Gyr.
An age difference greater than 1.0 Gyr is also obtained through the \citet{2008ApJ...687.1264M}
and \citet{2016A&A...594L...3L} calibrations.\\

\item[--] In fact, the value of $\Delta=0.24$ dex we found in the present work for most
recent years, has increased from the value $\Delta=0.22$ dex reported in \citep{2018MNRAS.476.2751F}\\

\item[--] The cycle amplitude of $\zeta^{2}$ Ret decreased notably, from $\Delta S_{MW}=0.0084$ to $\Delta S_{MW}=0.0045$
in the last cycle i.e. decreased by $\sim$47\%.
We caution that, until now, there is no clear agreement about the solar cycle behaviour during the MM state.
Initially, sunspot records suggest that the cycle was interrupted 
\citep[e.g.][]{1890MNRAS..50..251S,1976Sci...192.1189E,1994CAS....24.....W}.
However, different works using sunspots counts and cosmogenic isotopes, indicate a weaker but persistent cycle
\citep[e.g.][]{1993A&A...276..549R,1998SoPh..181..237B,2004SoPh..224..317M,2014SoPh..289.4701P,2015A&A...577A..71V}. 
In addition to the Sun, \citet{Shah_2018} showed that the G5V star HD 4915, a candidate to present a MM state,
also reveals a cycle still in progress with decreasing amplitude, similar to $\zeta^{2}$ Ret.\\

{\item[--] The current activity level of $\zeta^{2}$ Ret is very low ($\langle F_{HK}\rangle\sim 3\times  10^6$ erg cm$^{-2}$ s$^{-1}$).
This value is even lower within the statistical error than the theoretical basal level for this object ($F_{HK}= (3.10 \pm 0.07)\times  10^6$ erg cm$^{-2}$ s$^{-1}$). The basal value was determined by estimating the sum of the non-photospheric basal flux 
 $F'_{HK}$ for B-V $=$ 0.60 \citep{2013A&A...549A.117M},
and the photospheric contribution estimated using a synthetic spectra
calculated with SYNTHE and ATLAS9 model atmospheres \citep{1993KurCD..13.....K}.
The stellar parameters of $\zeta^{2}$ Ret were taken from the high-precision analysis
of \citet{2016A&A...588A..81S}.}

\end{itemize}

To find an unambiguous example of a MM candidate is a very difficult task,
which is due in part to the fact that a MM state in general is not totally clear,
and requires a large observational time span.
Up to now, there are only very few  MM candidates reported in the literature, being one of
them HD\,4915 \citep{Shah_2018}.
For the case of $\zeta^{2}$ Ret, we profit from the fact that this star belongs to
a binary system, which makes it a very valuable candidate. In fact, to our knowledge $\zeta^{2}$ Ret is the first MM candidate detected through the activity differences in a binary system, as suggested by DO98.

Finally, we strongly recommend the search for valuable MM candidates in those binary systems
with high activity differences. They can be used as good laboratories to address many questions related to solar/stellar MM,
which are yet subjected to intense scrutiny \citep[see][for more details]{Zolotova_2015,2015A&A...577A..71V,2019ApJ...886...18C}.

\begin{acknowledgements}
We warmly thank the anonymous referee for constructive comments that improved the paper. MF, MJA, RIB, NN, and PM acknowledge the financial support  
of PROJOVI/UNSJ, through the project 80020190300048SJ. Also, RIB, JA, and PM acknowledge the financial support from CONICET in the forms of
doctoral and post-doctoral fellowships. JYG acknowledges the support from CNPq.
\end{acknowledgements}

%
%

\bibliographystyle{aa}
\small
\bibliography{references}

\begin{appendix} 
\section{Observational data and derived Mount Wilson indexes for $\zeta$ Ret.}

\longtab[1]{
\begin{longtable}{clcllclcll}
\caption{Observational data and derived $\mathrm{S}_\mathrm{{MW}}$ indexes for the binary system $\zeta$ Ret.}\label{tabone}\\
\hline\hline
&              & HD\,20766 &      &                   &                            &                & HD\,20807 &               &           \\
\hline
Observing date & xJD & Instrument & Observing program & $\mathrm{S}_\mathrm{{MW}}$ & Observing date & xJD & Instrument & Observing program & $\mathrm{S}_\mathrm{{MW}}$ \\
\hline
\endfirsthead
\caption{continued.}\\
\hline\hline
&              & HD\,20766 &      &                   &                            &                &     HD\,20807 &           &               \\
\hline
Observing date & xJD & Instrument & Obs. program & $\mathrm{S}_\mathrm{{MW}}$ & Observing date & xJD & Instrument & Obs. program & $\mathrm{S}_\mathrm{{MW}}$ \\
\hline
\endhead
\hline
\endfoot
2000-08-12    & 1769       & REOSC       & 0800(B)         & 0.242     &    2002-12-05   & 2613.50   & UVES         & 266.D-5655(A)     & 0.176    \\
2002-08-01    & 2519       & REOSC       & 0802(C)         & 0.244     &    2002-12-05   & 2613.50   & UVES         & 266.D-5655(A)     & 0.176    \\
2002-11-21    & 2600       & REOSC       & 1102(A)         & 0.260     &    2002-12-05   & 2613.50   & UVES         & 266.D-5655(A)     & 0.177    \\
2003-11-06    & 2949.25    & HARPS       & 072.C-0513(B)   & 0.266     &    2002-12-05   & 2613.50   & UVES         & 266.D-5655(A)     & 0.177    \\ 
2003-11-06    & 2949.26    & HARPS       & 072.C-0513(B)   & 0.266     &    2003-10-27   & 2939.26   & HARPS        & 072.C-0488(E)     & 0.179    \\ 
2003-11-06    & 2949.26    & HARPS       & 072.C-0513(B)   & 0.266     &    2003-10-30   & 2942.23   & HARPS        & 072.C-0488(E)     & 0.178    \\ 
2003-11-06    & 2949.26    & HARPS       & 072.C-0513(B)   & 0.267     &    2003-11-06   & 2949.27   & HARPS        & 072.C-0513(B)     & 0.176    \\ 
2003-11-06    & 2949.26    & HARPS       & 072.C-0513(B)   & 0.268     &    2003-11-06   & 2949.27   & HARPS        & 072.C-0513(B)     & 0.176    \\
\end{longtable}
\tablefoot{Columns 2 and 7: Julian day xJD=JD-2\,450\,000. Columns 4 and 9: ID programs corresponding to ESO and REOSC observations.}
}

\end{appendix}

\end{document}